\documentclass[apl, superscriptaddress, preprint]{revtex4-1}
\usepackage{graphicx}% Include figure files
\usepackage{bm}% bold math

\usepackage{hyperref}% add hypertext capabilities
\hypersetup{
            colorlinks=true,
            linkcolor=blue,
            anchorcolor=blue,
            citecolor=blue
            }
\usepackage{changes}
\usepackage{amsmath, amssymb}

\begin{document}

%\preprint{APS/123-QED}
\title{Voltage-controlled significant spin wave Doppler shift in FM/FE heterojunction} 

%\author{Shaojie Hu}
%\email[]{shaojiehu@mail.xjtu.edu.cn}
%\affiliation{Center for Spintronics and Quantum Systems, State Key Laboratory for Mechanical Behavior of Materials, School of Materials Science and Engineering, Xi'an Jiaotong University, Xi'an, Shaanxi, 710049, China}

\author{Shaojie Hu}
%\email[]{hu.shaojie@phys.kyushu-u.ac.jp} 
\affiliation{Department of Physics, Kyushu University, 744 Motooka, Fukuoka, 819-0395, Japan}

\author{Kang Wang}
%\email[]{wangkang146503@163.com,3120102052@stu.xjtu.edu.cn}
\affiliation{Center for Spintronics and Quantum Systems, State Key Laboratory for Mechanical Behavior of Materials, School of Materials Science and Engineering, Xi'an Jiaotong University, Xi'an, Shaanxi, 710049, China}

\author{Tai Min}
%\email[]{tai.min@xjtu.edu.cn}
\affiliation{Center for Spintronics and Quantum Systems, State Key Laboratory for Mechanical Behavior of Materials, School of Materials Science and Engineering, Xi'an Jiaotong University, Xi'an, Shaanxi, 710049, China}

\author{Takashi Kimura}
%\email[]{t-kimu@phys.kyushu-u.ac.jp}
\affiliation{Department of Physics, Kyushu University, 744 Motooka, Fukuoka, 819-0395, Japan}

\date{\today}

\begin{abstract}
Efficient manipulation of spin waves (SWs) is considered as one of the promising means for encoding information with low power consumption in next generation spintronic devices. The SW Doppler shift is one important phenomenon by manipulating SWs propagation. Here, we predict an efficient way to control the SW Doppler shift by voltage control magnetic anisotropy boundary (MAB) movement in FM/FE heterojunction. From the micromagnetic simulation, we verified that the SW Doppler shift aligns well with our theoretical predictions in Fe/$\rm BaTiO_3$ heterostructure. The SW Doppler shift also shows ultra wide-band shift (over 5 GHz) property with voltage. Such efficient SW Doppler shift may provide one possible way to measure the Hawking radiation of an analogue black hole in the SW systems.

\end{abstract}
\maketitle
Magnonics, a burgeoning field rooted in the exploration of quantum magnetic dynamic phenomena, is primarily dedicated to the in-depth research and practical application of spin waves (SWs). These SWs hold promise for advanced information processing at the nanoscale, potentially revolutionizing how we store and compute data with ultra-low power consumption.\cite{neusser2009magnonics,2010Kruglyak, 2010Serga, 2011LENK,demokritov2012magnonics,barman20212021,pirro2021advances}
Efficiently controlling the spin wave propagation offers a potential alternative for encoding information beyond CMOS computing due to their absence of Joule heating during propagation, scalability to atomic dimensions, and diverse nonlinear and nonreciprocal phenomena.\cite{2005Kostylev, 2008Lee, 2014Klingler,2020Dieny}  Much like sound or light waves, SWs inherently exhibit wave properties, including the Doppler effect that denotes frequency shifts due to relative motion between the source and observer. Recent studies have highlighted numerous innovative SW Doppler shift occurrences. 
Stancil, \textit{et al.} reported the experimental observation of an inverse Doppler shift from left-handed dipolar SWs with a moving pick-up antenna, which is the intrinsic characteristic of left-handed, or backward SWs.\cite{2006Stancil} 
Vlaminck and Bailleul found the current-induced SW Doppler shift that related to the adiabatic spin-transfer torque, which can be used as a probe of spin-polarized transport in various metallic ferromagnets.\cite{2008Vlaminck} 
Meanwhile,  Yu, \textit{et al.} predicted the SW Doppler shift by magnon drag in magnetic insulators.\cite{2021Yu} 
In parallel,  Nakane and Kohno studied the current-induced SW Doppler shift in antiferromagnets.\cite{2021Jotaro}
So, researchers have hypothesized that it's possible to create magnonic black-hole and white-hole horizons using an efficient method involving spin current-driven SW Doppler shifts. 
This means that the SW Doppler shift might not only provide a mechanism for simulating gravitational phenomena, but also offer a way to study analogue gravity and even measure the Hawking radiation emitted by an analogous black hole.\cite{jannes2011hawking,roldan2017magnonic}
The intriguing nature of these phenomena has significantly heightened research interest in the SW Doppler shift, pushing scientists and scholars to delve deeper into its intricacies and potential applications.\par

So far, the preponderance of investigations into the SW Doppler shift has been concentrated on individual ferromagnetic materials that are modulated through electric current control.
Such a constrained approach inadvertently poses limitations on realizing the aspirational goal of ultra-low power consumption pivotal for the nuanced manipulation of SWs. Compounding this challenge, it has been observed that the SW Doppler shift frequencies are consistently far from the GHz benchmark. 
In the broader academic discourse, the magnetoelectric effect emerges as a promising avenue, widely regarded for its efficacy in directing the propagation of SWs. \cite{2018Balinskiy, 2018Sadovnikov,2018Rana,2019Rana,2021Qin,wang2022dual}
This is achieved with a commendable reduction in power consumption, especially when deployed within multiferroic materials. 
In the pantheon of multiferroic systems, the artificial ferromagnetic/ferroelectric (FM/FE) heterojunction stands as an archetypal representative.\cite{sahoo2007ferroelectric,venkataiah2012strain,Lahtinen2012,venkataiah2013strain,gorige2017magnetization,Yamada2021Electric,Usami2021,Fujii2022Giant,hu2023efficient} Within this context, we present a pronounced SW Doppler shift attributed to the motion of the magnetic anisotropy boundary (MAB) inherent in the FM/FE heterojunction, a process which is deftly modulated by voltage.

 \begin{figure}[htb]
	\centering
	\includegraphics[width=5.5in]{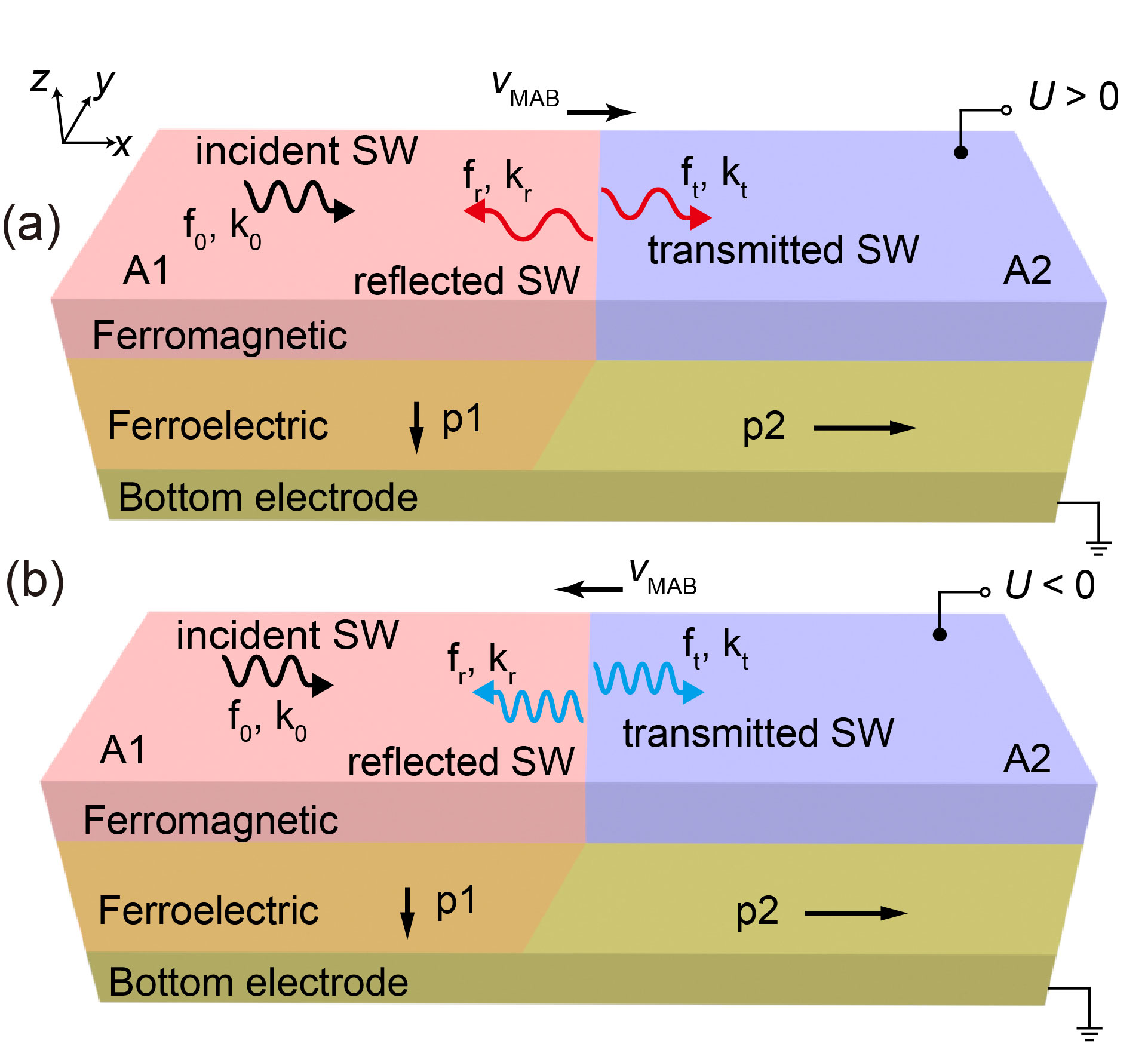}	
	\caption{ The schematic of SW Doppler shift induced by the voltage-controlled MAB movement in FM/FE heterojunction. The out-of-plane and in-plane polarization domains in the ferroelectric layer transfer different strains to the ferromagnetic layer. The ferromagnetic layer exhibits different anisotropy regions owing to inverse magnetostriction. (a) A case of SW Doppler red shift caused by a positive voltage. (b) A case of SW Doppler blue shift caused by a negative voltage. 
	}
	\label{figure1}
\end{figure}

Figure \ref{figure1} provides a schematic representation of SW Doppler shift precipitated by the voltage-controlled MAB movement in FM/FE heterojunction.  Within this configuration, the ferromagnetic layer, positioned atop a ferroelectric substrate with both in-plane and out-of-plane polarized domains, manifests two distinct magnetic anisotropy regions: A1 and A2.  Varied magnetic anisotropies result from the strain-induced inverse magnetostriction effect across distinct polarized domains.\cite{2012Lahtinen} 
The MAB separating the two magnetic anisotropy regions is pinned on the ferroelectric domain wall and parallel to \emph{y} axis. A positive (negative) voltage-controlled perpendicular electric field can drive the motion of the ferroelectric domain wall along \emph{x} (\emph{-x}) axis as well as the MAB.\cite{2015Franke} 
The frequencies of reflected and transmitted SWs will show the red or blue shift by moving MAB. This letter presents an analytical model of the SW Doppler shift, which is further validated through magnetic simulation.
 
For simplicity, we assume that the MAB moves at a constant velocity $v_\mathrm{MAB}$. The incident SW propagates along \emph{x} axis and perpendicular to the MAB.
In the observer's system S which is stationary with respect to the lab frame, the general dispersion relations of SWs in A1 and A2 are described as different functions $\rm f_1 = F_1(\mathbf{k})$ and  $\rm f_2 = F_2(\mathbf{k})$ with wave vector $\mathbf{k}$.  
In the moving system S' which is stationary with respect to the MAB, the dispersion functions  $\rm f_1' = F_1'(\mathbf{k})$ and  $\rm f_2' = F_2'(\mathbf{k}$) with vector $\mathbf{k'}$ in A1 and A2  are given as follows by Lorentz transformation:
\begin{equation}
    \begin{aligned}
        f'_{i=1,2} &= F'_{i=1,2}(\mathbf{k'})=\Gamma \left (F_{i=1,2}(\mathbf{k'}) - \frac{\mathbf{k'}\cdot \mathbf{v}_\mathrm{MAB}}{2\pi}\right ),\  \mathbf{k'}_{i=1,2} &= \Gamma \left (\mathbf{k}_{i=1,2} - \frac{2\pi f_{i=1,2}\mathbf{v}_\mathrm{MAB}}{c^2}\right )
    \end{aligned}
\end{equation}

%\begin{equation}
%    \begin{aligned}
%        \mathbf{k'}_{i=1,2} &= \Gamma \left (\mathbf{k}_{i=1,2} - \frac{2\pi f_{i=1,2}\mathbf{v}_\mathrm{MAB}}{c^2}\right ) \\
%    \end{aligned}
%\end{equation}
where $\Gamma = \left (1 - v_\mathrm{MAB}^2/s^2\right )^{-\frac{1}{2}}$, $s$ is the maximal group velocity of spin waves.\cite{kim2014propulsion} Under the limitation of $v_\mathrm{MAB}$ $\ll$ \ $s$, the $f'$ and $\mathbf{k}'$ can be written as the following:
\begin{equation}
    \begin{aligned}
        f'_{i=1,2} &= F'_{i=1,2}(\mathbf{k'}) &\approx F_{i=1,2}(\mathbf{k'}) - \frac{\mathbf{k'}\cdot \mathbf{v}_\mathrm{MAB}}{2\pi}, \   \mathbf{k}'_{i=1,2} &\approx \mathbf{k}_{i=1,2} \\
    \end{aligned}
    \label{mf}
\end{equation}

The frequencies and wave vectors of incident, reflected and transmitted SWs in system S and S' are defined as: $f_0$, $\mathbf{k_0}$, $f_\mathrm{r}$, $\mathbf{k}_\mathrm{r}$, $f_\mathrm{t}$, $\mathbf{k}_\mathrm{t}$, $f'_0$, $\mathbf{k}'_0$, $f'_\mathrm{r}$, $\mathbf{k}'_\mathrm{r}$, $f'_\mathrm{t}$ and $\mathbf{k}'_\mathrm{t}$, respectively.
If the incident SW from the A1 region propagates to A2 shown in Fig.\ref{figure1}, the frequencies of reflected and transmitted SWs in system S can be derived as follows:
\begin{equation}
    \begin{aligned}
    f_\mathrm{r} &= \Gamma \left (f'_\mathrm{r} + \frac{\mathbf{k}'_\mathrm{r}\cdot\mathbf{v}_\mathrm{MAB}}{2\pi}\right) \\
    &\approx f_0 - (\mathbf{k}_0 - \mathbf{k}'_\mathrm{r})\cdot\frac{\mathbf{v}_\mathrm{MAB}}{2\pi}
    \end{aligned}
    \label{reflected_f}
\end{equation}

\begin{equation}
    \begin{aligned}
    f_\mathrm{t} &= \Gamma \left(f'_\mathrm{t} + \frac{\mathbf{k}'_\mathrm{t}\cdot\mathbf{v}_\mathrm{MAB}}{2\pi}\right) \\
    &\approx f_0 - (\mathbf{k}_0-\mathbf{k}'_\mathrm{t})\cdot\frac{\mathbf{v}_\mathrm{MAB}}{2\pi}
    \end{aligned}
    \label{transmitted_f}
\end{equation}
where, incident frequency $f_0$ is known, $\mathbf{k}_0$ can be obtained from the dispersion relation of SWs in the incident region. $\mathbf{k}'_\mathrm{r}$ and $\mathbf{k}'_\mathrm{t}$ can be obtained from Eq. (\ref{mf}), respectively. 
 %The detailed derivation is in the Supplementary Material.
  So the frequency shift of SW Doppler effect is dominated by the second terms of Eq. (\ref{reflected_f}) and Eq. (\ref{transmitted_f}), which not only relate to the velocity of MAB but also the deviation of the incident and reflected (or transmitted) wave vectors
  determined by different magnetic anisotropy. 
 
\begin{figure}[htb]
	\centering
	\includegraphics[width=6in]{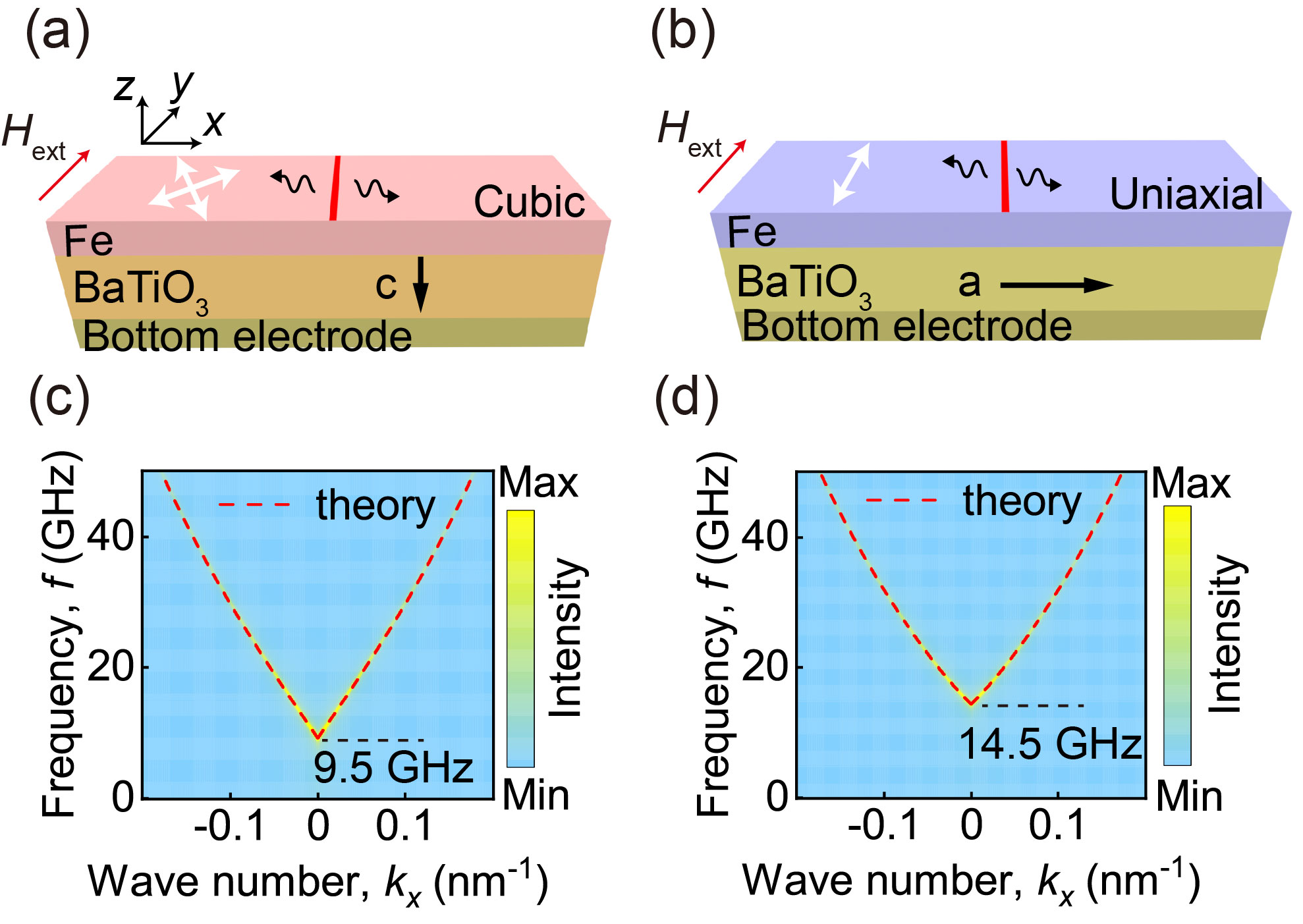}	
	\caption{  
	(a) The cubic anisotropy is induced in $\mathrm{Fe}$ layer on ferroelectric $\mathrm{BaTiO_3}$ c-domain. The two white double-head arrows represent easy axes of cubic anisotropy along with $45^{\circ}$ to \emph{y} axis. 
    (b) The uniaxial anisotropy is induced in $\mathrm{Fe}$ layer on ferroelectric $\mathrm{BaTiO_3}$ a-domain. The white double-head arrows represent the easy axis of uniaxial anisotropy along \emph{y} axis. The image plots of the dispersion relation of SWs in cubic anisotropy region (c) and uniaxial anisotropy region (d) by micromagnetic simulation. The dashed lines correspond to the theory calculation data. 
	}
	\label{figure2}
\end{figure}
To validate such SW Doppler shift, we conducted the micromagnetic simulation in a well-build artificial multiferroic system  Fe/$\mathrm{BaTiO_3}$ \cite{2006Duan, 2014Radaelli} by using the GPU-accelerated micromagnetic software Mumax3\cite{2014Vansteenkiste}.
Figure \ref{figure2}(a) and (b) show the Fe film on ferroelectric $\mathrm{BaTiO_3}$ substrate with in-plane and out-of-plane polarization domains exhibits uniaxial and cubic magnetic anisotropy.\cite{2012Lahtinen} 
In the simulation, a $6000 \times 400 \times 2\ \mathrm{nm}^3$ Fe film is discretized using $3000 \times 200 \times 1$ finite difference cells. The periodic boundary condition is used along \emph{y} axis to avoid the boundary effect. 
The simulation parameters used are as follows\cite{2015Franke,2021Qin}: saturation magnetization $M_\mathrm{s} = 1.7 \times 10^6$ A/m, exchange constant $A_\mathrm{ex} = 2.1 \times 10^{-11}$ J/m, Gilbert damping $\alpha$ = 0.003, cubic anisotropy constant $K_\mathrm{c} = 4.4\times 10^4\ \mathrm{J/m^3}$ and uniaxial anisotropy constant $K_\mathrm{u} = 1.5 \times 10^4\ \mathrm{J/m^3}$. 
To prevent SW reflection at both ends, $\alpha$ is gradually increased from 0.003 to 0.033 at both end regions of the Fe film ($-3000$ nm \textless \ \emph{x} \textless \ $-2200$ nm, $2200$ nm \textless \ \emph{x} \textless \ $3000$ nm).  
An external magnetic field $\mu _0 H_\mathrm{ext} = 100\ \mathrm{mT}$ is applied to magnetize the Fe film along \emph{y} axis. 
To obtain the dispersion relation of SWs, we perform two dimensional Fourier transform\cite{2012Kumar,2013Venkat} on $m_x(x, t)$ in response to a sinc-based exciting field $ \mathrm{h}_0\mathrm{sinc}(2\pi f_\mathrm{c}(t-t_0))\hat{e}_x$ with $\mu _0h_0 = 10\ \mathrm{mT}$, cutoff frequency $f_c = 50\ \mathrm{GHz}$ and $t_0 = 5$ ns, at the center section ($6 \times 400 \times 2\ \mathrm{nm}^3$) of Fe film. The dispersion relations are shown in Fig. \ref{figure2}(c) and (d) for cubic and uniaxial regions, respectively.
The red dashed lines represent the theoretical calculations from the dispersion relation equations of Damon–Eshbach (DE) SWs in cubic and uniaxial anisotropy regions as follow\cite{1986Kalinikos,wang2022dual}:

\begin{equation}
    f_{cu} = F_{cu}(k) = \frac{\gamma \mu _0}{2\pi}\sqrt{\left(H_{\mathrm{ext}}-\frac{2K_{\mathrm{c}}}{\mu _0M_{\mathrm{s}}} + \frac{2A_{\mathrm{ex}}}{\mu_0 M_{\mathrm{s}}}k^2\right ) \left (H_{\mathrm{ext}}+\frac{K_{\mathrm{c}}}{\mu _0M_{\mathrm{s}}} + M_{\mathrm{s}} + \frac{2A_{\mathrm{ex}}}{\mu_0 M_{\mathrm{s}}}k^2\right ) + M^2_{\mathrm{s}}F}
    \label{fKc}
\end{equation}

\begin{equation}
    f_{ku} = F_{ku}(k)=\frac{\gamma \mu _0}{2\pi}\sqrt{\left(H_{\mathrm{ext}}+\frac{2K_{\mathrm{u}}}{\mu _0M_{\mathrm{s}}} + \frac{2A_{\mathrm{ex}}}{\mu_0 M_{\mathrm{s}}}k^2\right) \left(H_{\mathrm{ext}}+\frac{2K_{\mathrm{u}}}{\mu _0M_{\mathrm{s}}} + M_{\mathrm{s}} + \frac{2A_{\mathrm{ex}}}{\mu_0 M_{\mathrm{s}}}k^2\right) + M^2_{\mathrm{s}}F}
      \label{fKu}
\end{equation}
where, $\gamma$ is gyromagnetic ratio, $\mu _0$ is vacuum permeability, $F = \frac{1-\mathrm{e}^{-\lvert k \rvert d}} {\lvert k \rvert d} (1-\frac{1-\mathrm{e}^{-\lvert k \rvert d}}{\lvert k\rvert d})$, $d$ is the thickness of Fe film. Obviously, the variance in anisotropy constants between uniaxial and cubic regions yields distinct dispersion relations.

\begin{figure}[htb]
	\centering
	\includegraphics[width=6.3in]{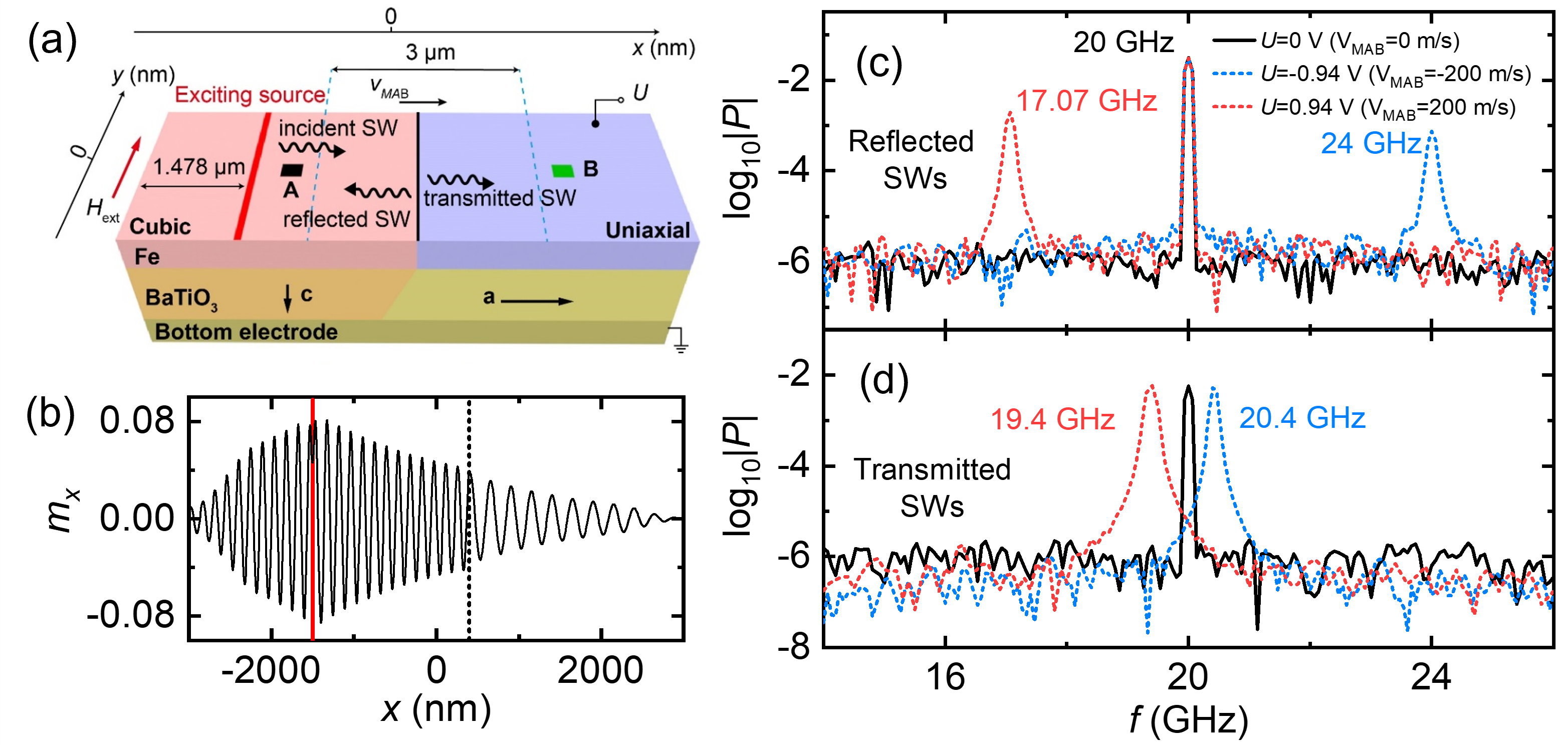}	
	\caption{(a) The schematic of reflected and transmitted SWs by moving MAB with voltage. The MAB moves along \emph{x} axis between $-1100$ nm and 1900 nm marked by the blue dash lines. The black square (A) represents the cell at $-1202$ nm \textless \ \emph{x} \textless \ $-1200$ nm, $-2$ nm \textless \ \emph{y} \textless \ $0$ nm, where $m_x(t)$ is recorded as the incident and reflected SWs. The green square (B) represents the cell at $1998$ nm \textless \ \emph{x} \textless \ $2000$ nm, $-2$ nm \textless \ \emph{y} \textless \ $0$ nm, where $m_x(t)$ is recorded as the transmitted SW. (b) The snapshot of $m_x$ taken at $y = -1$ nm with the static MAB at \emph{x} = 400 nm. The red line is the position of exciting source and the black dashed line is the position of MAB. (c) The spectra of reflected and incident SWs when $v_\mathrm{MAB} = -200$, 0 and 200 m/s. (d) The spectra of transmitted SW when $v_\mathrm{MAB} = -200$, 0 and 200 m/s.
	}
	\label{figure3}
\end{figure}

We then simulate the reflection and transmission of SWs by moving MAB, as depicted in Fig. \ref{figure3}(a). 
%and we neglect it in simulation. 
The exciting source of SWs is 1.478 $\mathrm{\mu m}$ from the left end with a size of $42 \times 400 \times 2\ \mathrm{nm}^3$. 
The SW is excited by a consistent Sine function field $ \mu _0\mathrm{h}_0\mathrm{sin}(2\pi f_\mathrm{0}t)\hat{e}_x$ ($\rm \mu _0h_0 = 10\ \mathrm{mT}$,  $f_0 = 20\ \mathrm{GHz}$) under the external magnetic field $\mu _0 H_\mathrm{ext} = 100\ \mathrm{mT}$.
The excited SW propagates in Fe film along \emph{x} axis perpendicular to the MAB. 
The width of MAB same as the pinned ferroelectric domain wall (2 - 5 nm \cite{1992Zhang, 2006Hlinka, 2006ZhangQingsong}) is far smaller than the wavelength of SWs and we neglect it in simulation.
The MAB moves along \emph{x} axis between \emph{x} = $-1100$ nm and \emph{x} = 1900 nm marked by the blue dash lines in Fig. \ref{figure3}(a). 
It has been confirmed the velocity of ferroelectric domain wall driven by electric field can be up to 1000 m/s in $\rm BaTiO_3$.\cite{1963Stadler, 2017Boddu}
And, the velocity of MAB ($v_\mathrm{MAB}$) could be calculated by Miller-Weinreich theory  as following equation\cite{1963Stadler,tagantsev2010domains}:
\begin{equation}
    v_\mathrm{MAB}(U) = v_\infty \sum_{n=1}^{\infty}n\mathrm{e}^{-(\delta/(U/D))n^{\frac{3}{2}}}
    \label{uvmab}
\end{equation}
where $U$ is the voltage to provide perpendicular electric field in $\rm BaTiO_3$ , $v_\infty$ = 5 m/s, $\delta$ = 4 kV/cm, the thickness of $\rm BaTiO_3$ substrate $D$ = 100 nm. So, a significant velocity of MAB could be realized by a much lower voltage.

\begin{figure}[htb]
	\centering
	\includegraphics[width=4.5in]{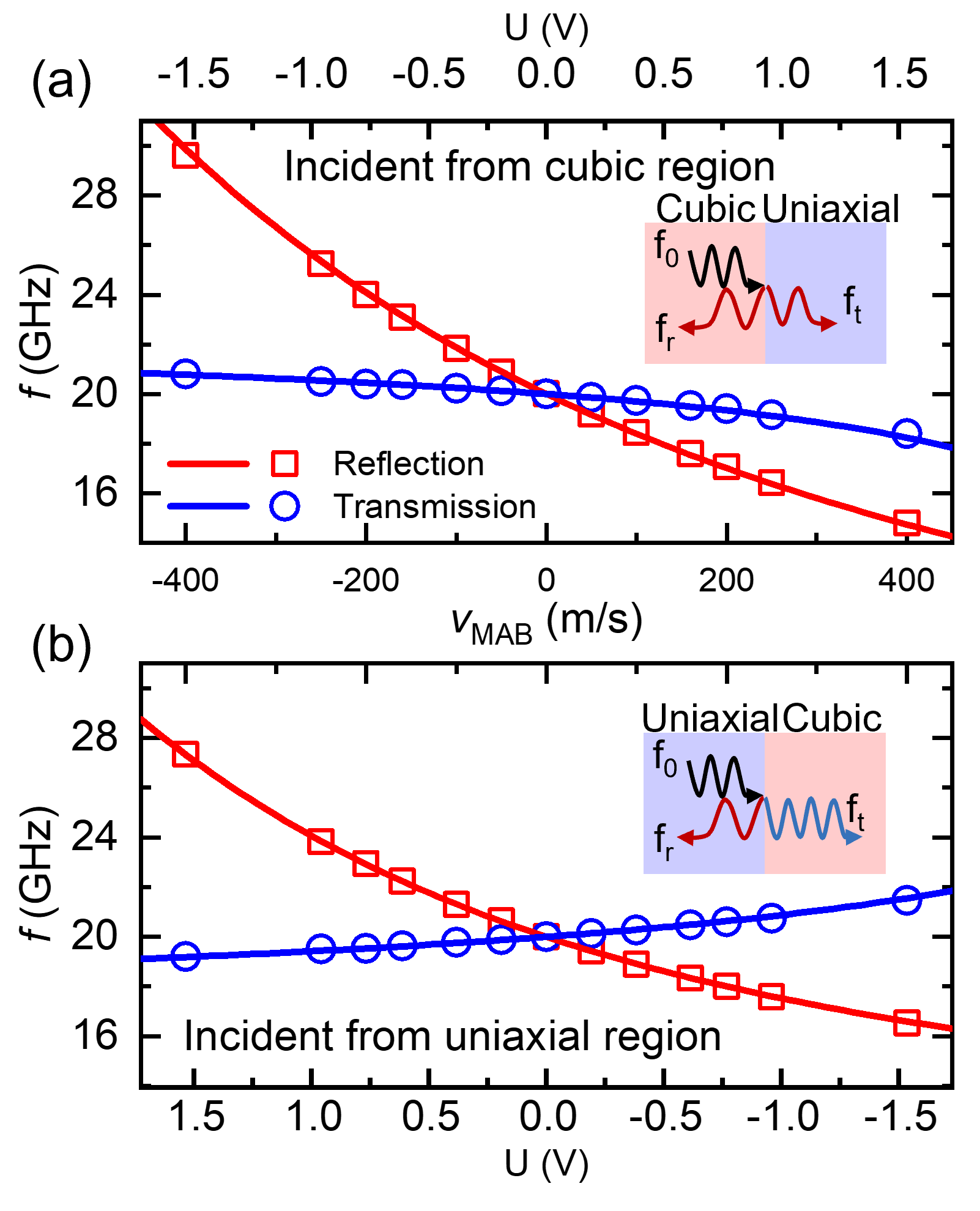}	
	\caption{ The frequencies of reflected and transmitted SWs as a function $v_\mathrm{MAB}$ (voltage) in Fe/$\rm BiTiO_3$ at the exciting frequency of 20 GHz.
 (a) The SWs incident from cubic to uniaxial region. The positive voltage provides the positive velocity of MAB.  (b) The SWs incident from uniaxial to the cubic regions. The negative voltage provides the positive velocity of MAB.
 The symbols are corresponding to the simulation results. The solid lines correspond to the theoretical calculation.
	}
	\label{figure4}
\end{figure}

In the simulation, we implement MAB moving at velocity $v_\mathrm{MAB}$ by shifting it over one discretization cell ($\delta x = 2$ nm) every time interval $\delta t = \delta x/v_\mathrm{MAB}$.\cite{2015Franke, 2016Wiele} 
The $m_x$ in the cell of $-1202$ nm \textless \ \emph{x} \textless \ $-1200$ nm, $-2$ nm \textless \ \emph{y} \textless \ $0$ nm is recorded as incident and reflected SWs, while the $m_x$ in the cell of $1998$ nm \textless \ \emph{x} \textless \ $2000$ nm, $-2$ nm \textless \ \emph{y} \textless \ $0$ nm is recorded as transmitted SW, marked by the black and green squares A, B in Fig. \ref{figure3}(a), respectively. 
Figure \ref{figure3}(b) shows the snapshot of $m_x$ taken at \emph{y} = $-1$ nm with static MAB at \emph{x} = 400 nm. The black dashed line indicates the position of MAB. Clearly, the SWs have different wavelengths in cubic and uniaxial regions.
To accurately determine all the related frequencies of the SWs, we perform a fast Fourier transform on $m_x(t)$ at A and B to obtain the spectra of reflected and transmitted SWs. 
Figure \ref{figure3}(c) shows the spectra of the incident and reflected SWs at various $v_\mathrm{MAB}$. 
At $v_\mathrm{MAB}$ = 0 m/s, a single 20 GHz peak in the spectrum (black solid curve) indicates the reflected SW  frequency matches exciting frequency. 
However, for  $v_\mathrm{MAB}$ = 200 m/s,  the spectrum (red dashed curve) displays peaks at 20 GHz and 17.07 GHz. The latter peak, the reflected SW, shows a significant red shift of the Doppler effect. 
With a negative voltage setting $v_\mathrm{MAB}$ = $-200$ m/s, 
reflected SW frequency blue shifts to 24 GHz. 
Figure \ref{figure3}(d) shows the spectra of transmitted SW for various MAB velocities. They all have a single peak due to the solitary transmitted spin wave in the B.  
The transmitted SW frequencies display a Doppler red shift to 19.4 GHz and blue shift to 20.4 GHz at $v_\mathrm{MAB}$ = 200 and $-200$ m/s respectively. Notably, these Doppler shifts are much lower than those of the reflected SWs, which hold opposing wave vectors to the incident SW.

To comprehensively analyze the Doppler shift, we simulate the SW Doppler shift across a broader range of velocities (voltages). 
Figure \ref{figure4}(a) plots both simulated and theoretical Doppler shifts of reflected and transmitted SWs' frequencies at the exciting frequency of 20 GHz in the cubic region. 
The simulation aligns closely with theoretical predictions. The frequencies of reflected and transmitted SWs show a monotonic shift with the velocity (voltage). This also means all the reflected and transmitted SWs show a red shift when the MAB is far away from the source of the SW. Meanwhile, the SWs show a blue shift when the MAB is close to the SW source. 
In addition, the SW Doppler shift can exceed several GHz, nearly two orders greater than the electric current control methods.  
It's also interesting the transmission SW shows a different velocity-dependent tendency when the incident SW from uniaxial region to cubic region shown in Fig. \ref{figure4}(b). Here, the transmitted SWs show a blue shift when the MAB is far away from the SW source, and a red shift when the MAB is close to the source. Such unusual Doppler shift comes from the negative value of  $\mathbf{k}_0-\mathbf{k_t}'$ based on the Eq.\ref{transmitted_f}. These results also indicate that the type of transmitted SW Doppler shift is influenced not only by the MAB's movement direction but also by the deviation between the two different regions. 
\par
In summary, we predict an efficient way to control the SW Doppler shift by voltage control MAB movement in FM/FE heterojunction. The SW Doppler shift could be modulated in an ultra-wide-band range with voltage. The micromagnetic simulation results align closely with theoretical predictions. Such a significant SW Doppler shift may provide one possible system to do analogue gravity and measure the Hawking radiation of an analogue black hole. On the other hand, the SW Doppler shift could be employed as an accurate probe of the ferroelectric domain wall motion controlled by electric field.

\begin{acknowledgments}
%This work is partially supported by National Key Research Program of China (Grant No. 2017YFA0206200), 
This work is partially supported by National JSPS Program for Grant-in-Aid for Scientific Research (S)(21H05021), and Challenging Exploratory Research (17H06227) and JST CREST (JPMJCR18J1). Natural Science Foundation of Shaanxi Province (2021JM-022).
\end{acknowledgments}
\bibliographystyle{unsrt}
\bibliography{Bib}

\end{document}